\PassOptionsToPackage{dvipsnames}{xcolor}
\documentclass[11pt,letterpaper]{article}
\usepackage[margin=1in]{geometry}
\usepackage{setspace}

\usepackage{tikz}
\usetikzlibrary{arrows}
\usepackage{filecontents}

\usepackage{mathtools}

\usepackage{float}
\usepackage{caption, subcaption}
\usepackage{graphicx}
\usepackage{amsmath,amsthm,amssymb,amsfonts}
\usepackage{algorithm}
\usepackage{listings}
\usepackage[colorlinks,linkcolor=blue]{hyperref}
\usepackage{url}
\usepackage{cleveref}

\usepackage{colortbl}

\usepackage{pdfpages}

\newcommand{\hlcolor}{Yellow!35}
\newcommand{\hlcolorTwo}{LimeGreen!35}
\makeatletter
\newenvironment{btHighlight}[1][]
{\begingroup\tikzset{bt@Highlight@par/.style={#1}}\begin{lrbox}{\@tempboxa}}
{\end{lrbox}\bt@HL@box[bt@Highlight@par]{\@tempboxa}\endgroup}

\newcommand\btHL[1][]{%
  \begin{btHighlight}[#1]\bgroup\aftergroup\bt@HL@endenv%
}
\def\bt@HL@endenv{%
  \end{btHighlight}%
  \egroup
}
\newcommand{\bt@HL@box}[2][]{%
  \tikz[#1]{%
    \pgfpathrectangle{\pgfpoint{1pt}{0pt}}{\pgfpoint{\wd #2}{\ht #2}}%
    \pgfusepath{use as bounding box}%
    \node[anchor=base west, fill=\hlcolor,outer sep=0pt,inner xsep=1pt, inner ysep=0pt, rounded corners=2pt, minimum height=\ht\strutbox+2pt,#1]{\raisebox{1pt}{\strut}\strut\usebox{#2}};
  }%
}

\newenvironment{btHighlightTwo}[1][]
{\begingroup\tikzset{bt@HighlightTwo@par/.style={#1}}\begin{lrbox}{\@tempboxa}}
{\end{lrbox}\bt@HLTwo@box[bt@HighlightTwo@par]{\@tempboxa}\endgroup}

\newcommand\btHLTwo[1][]{%
  \begin{btHighlightTwo}[#1]\bgroup\aftergroup\bt@HLTwo@endenv%
}
\def\bt@HLTwo@endenv{%
  \end{btHighlightTwo}%
  \egroup
}

\newcommand{\bt@HLTwo@box}[2][]{%
  \tikz[#1]{%
    \pgfpathrectangle{\pgfpoint{1pt}{0pt}}{\pgfpoint{\wd #2}{\ht #2}}%
    \pgfusepath{use as bounding box}%
    \node[anchor=base west, fill=\hlcolorTwo,outer sep=0pt,inner xsep=1pt, inner ysep=0pt, rounded corners=2pt, minimum height=\ht\strutbox+2pt,#1]{\raisebox{1pt}{\strut}\strut\usebox{#2}};
  }%
}

\newcommand{\code}[1]{\texttt{#1}}

\author{Taras Skazhenik$^1$ \and Nikolai Korobenikov$^1$ \and Andrei Churbanov$^2$ \and Anton Malakhov$^2$ \and Vitaly Aksenov$^1$}

\date{%
    $^1$ITMO University, Russia\\%
    $^2$Huawei, Russia\\[2ex]
}

\title{Basic Lock Algorithms in Lightweight Thread Environments}

\begin{document}

\singlespacing

\maketitle

\begin{abstract}
Traditionally, multithreaded data structures have been designed for access by the threads of Operating Systems (OS). However, implementations for access by programmable alternatives known as lightweight threads (also referred to as asynchronous calls or coroutines) have not been thoroughly studied. The main advantage of lightweight threads is their significantly lower overhead during launch and context switching. However, this comes at a cost: to achieve proper parallelism, context switches must be manually invoked in the code; without these switches, new lightweight threads will never be executed.

In this paper, we focus on the simplest multithreaded data structure: a mutex (also known as a lock). We demonstrate that original implementations for OS threads cannot be used effectively in this new context due to the potential for deadlocks. Furthermore, correctness is not the only concern. In certain languages, such as C\code{++}, there are various lightweight thread libraries, each with different implementations and interfaces, which necessitate distinct lock implementations.

In this work, we present a modification of TTAS and MCS locks for the use from lightweight threads and demonstrate that the two context switch mechanisms of lightweight threads, yielding and sleeping, are crucial. However, the performance of TTAS and MCS may differ significantly depending on the settings. If one wants to have a lock that works well for any library, we suggest using the cohort lock, which strikes a balance between MCS and TTAS by utilizing several MCS queues with a common TTAS.

\vspace{0.5cm}

\end{abstract}

\newpage

\section{Introduction}
\label{sec:introduction}
In 1965, Gordon Moore~\cite{Moore1965} observed that the number of transistors on integrated circuits doubles roughly every two years, driving exponential performance growth. This trend persisted for decades, driven by advances in semiconductor technology and microarchitectural optimizations. By the early 2000s, however, transistor scaling alone had become economically and physically infeasible \cite{Sutter2005/FreeLauncIsOver}, prompting a shift toward multicore processors and multiprocessor systems.

Traditionally, such systems relied on Operating System (OS) threads, which provide parallelism but incur substantial overhead from context switching and scheduler interaction. To reduce these costs, user-space lightweight threads emerged. Lightweight threads can refer to a variety of abstractions, but in general, they fall into two categories: stackful and stackless coroutines. Stackless coroutines rely on compiler-generated state machines, can only suspend at predefined locations, and cannot be run on a different OS thread, reducing the possibility for concurrency. Stackful coroutines, in contrast, maintain their own execution stack, meaning that they resemble OS threads in terms of functionality and, thus, they are also known as lightweight threads. However, compared to OS threads, they benefit from substantially lower context-switching costs due to user-space scheduling that bypasses the kernel. On the other hand, they request a specific manual management approach, utilizing manually placed context switches. In this work, we use the term ``lightweight threads'' exclusively for stackful coroutines, because they serve as a full replacement for OS threads, unlike stackless coroutines, which are better suited to asynchronous programming models than to classical concurrency.

Now, we briefly explain how systems of lightweight threads work. The user first creates OS threads being \texttt{carrier threads}, typically in a one-to-one ratio with the number of available CPU cores. Then, the system creates pools of lightweight threads for these carrier threads, where lightweight threads reside while waiting for execution. Finally, a scheduler acts as the coordinating component between these abstractions. The scheduler monitors the system state, extracts threads from the pools, dispatches them for execution on carrier threads, and returns those that have performed a context switch back into the pools. The parameters above, such as the number of pools, how to share them to carrier threads, the ordering of threads within a pool, and the scheduling algorithms, are limited only by the capabilities of the library and the developer's design choices. For example, one may employ a single global pool for the entire system or maintain a separate pool per a carrier thread. The ordering within a pool may follow FIFO, LIFO, or other sophisticated priority. While library-provided schedulers are typically based on well-known concepts such as round-robin or work-stealing, they often allow custom logic to be implemented.

Interaction with lightweight threads is a broad topic. In this work, we focus on one of the fundamental challenges in multithreaded programming: ensuring mutual exclusion when accessing critical sections of code. This problem is typically addressed through the use of a shared object that exists in one of two states: locked or unlocked~\cite{Tanenbaum2015/ModernOS}. This object is typically referred to as a mutex or lock. Research papers have presented numerous implementations~\cite{Rudolph1984/TTAS,MellorCrummey1991/MCS,Craig1993/CLH,Chabbi2016/HMCS,Dice2019/CNA,Dice2012/LockCohorting,Dice2025/ReciprocatingLocks}. They have been evaluated for OS threads and frequently employ waiting strategies incompatible with lightweight threading. Classical locks were designed for preemptive multitasking, where the OS controls thread scheduling.
In cooperative environments, one may yield or spawn a new thread inside the critical section.
Such behaviour is not uncommon~\cite{OpenBLASIssue2255} and, for example, can happen while using complicated functions from high-performance mathematical libraries: one can call from the critical section a function that uses parallelism inside it.
Unfortunately, if we use existing lock implementations, a thread holding a lock may yield execution while others spin indefinitely, causing a deadlock.
Thus, one has to adapt them for lightweight threads. Please note that the performance characteristics of the new locks may differ those of the original ones.

When analyzing the two most popular libraries with lightweight threading, Argobots \cite{Seo2017/Argobots, Castello2020/AnalysisLTLibs} and Boost Fibers \cite{BoostFiber}, we found two main mechanisms for operating with lightweight threads: yielding and suspending/resuming. In this work, we adapt two basic lock algorithms, TTAS~\cite{Rudolph1984/TTAS} and MCS~\cite{MellorCrummey1991/MCS}, to work with lightweight threads and demonstrate that both mechanisms, yielding and suspending, are essential and should be used depending on the library and machine settings. Additionally, the relative performance of TTAS and MCS can differ a lot: in some settings, TTAS significantly outperforms MCS, while in some others, MCS significantly outperforms TTAS. If one wants to use a lock that works well independently of the library, we suggest a version of a cohort lock~\cite{Dice2012/LockCohorting} that balances between TTAS and MCS: it uses several MCS queues with one common TTAS.

\paragraph{Roadmap. } In Section~\ref{sec:related}, we discuss the related work. In Section~\ref{sec:common-adpating-method}, we explain how to adapt MCS and TTAS to work with lightweight threads and propose a version of a cohort lock that balances between MCS and TTAS. In Section~\ref{sec:settings}, we introduce the workloads on which we run the new lock implementations. In Section~\ref{sec:evaluations}, we discuss the results of the evaluation. Finally, we conclude in Section~\ref{sec:conclusion}.

\section{Related Work}
\label{sec:related}
Unfortunately, we are unaware of any works that study the locks in the context of non-preemptive lightweight threads, but we briefly review the established concepts and alternative methods that achieve comparable results through different techniques.

\paragraph{Preemptive lightweight threads.} It is worth emphasizing that there are two principal ways to use lightweight threads. The first is to explicitly adapt application code, which will be our focus in this study. The second is to make non-preemptive threads preemptive through different techniques~\cite{Shiina2021/PreemptiveLightThreads, Madsen2021/PreemptiveCoroutines, Boucher2020/LwPreemptibleFunc, Mollison2013/UserspaceLibWithPreemptiveThreads}. One recent study \cite{Shiina2021/PreemptiveLightThreads} introduced an approach for implicit lightweight thread preemption with minimal overhead and evaluates them on popular systems.
However, the authors also note that non-preemptive lightweight threads with explicitly annotated context-switch points achieve the highest efficiency, although their adoption requires substantial code refactoring, which is rarely feasible in large-scale projects.
The difficulties of using non-preemptive lightweight threads are common across languages. For instance, Go made goroutines asynchronously preemptible in version 1.14 to avoid deadlocks \cite{Golang114, Donovan2015/Go}. Similarly, Java faced challenges with pinning virtual threads to OS threads in synchronized methods until version 24~\cite{JavaJep491}. However, busy-waiting loops are never preempted by the Java scheduler, and lightweight thread switching occurs only on blocking operations. Therefore, lock implementations in the standard library rely on thread parking.

\paragraph{Locks overview.} Various approaches to mutual exclusion have been extensively studied and remain an active research topic to this day. The simplest implementations utilize constructs based on one or several atomic variables, such as Test-Test-And-Set (TTAS)~\cite{Rudolph1984/TTAS} or Ticket Lock \cite{MellorCrummey1991/MCS}. More sophisticated approaches employ request queues, exemplified by MCS~\cite{MellorCrummey1991/MCS} and CLH~\cite{Craig1993/CLH}. Finally, recently proposed implementations like HMCS~\cite{Chabbi2016/HMCS} and CNA~\cite{Dice2019/CNA} leverage complex hierarchical structures to optimize performance on NUMA architecture processors \cite{Lameter2013/NUMA}. Most studies conducted experiments on x86 architecture processors \cite{Shanley2010/x86}, though some \cite{Oberhauser2021/HMCSArm} include results from ARM architecture \cite{Goodacre2005/ARM} processors, which have gained significant traction in recent years. We also highlight a method for transforming general-purpose mutexes into NUMA-aware ones, known as lock cohorting \cite{Dice2012/LockCohorting}. This approach is actively used in our work to combine TTAS as the outer lock with multiple MCS queues serving as the slow path. Research on different lock designs continues to this day, for instance, a new perspective was recently introduced in the work on Reciprocating Locks \cite{Dice2025/ReciprocatingLocks}. 

In the context of the programming languages mentioned above, lightweight threads employ similar lock designs regardless of their scheduling strategy. In Go, the locking mechanism uses a short-path and slow-path approach: a thread first attempts to acquire the lock using a single Compare-and-Swap (CAS) operation; upon failure, it spins and retries several times. If these attempts do not succeed, the thread is enqueued and suspended. To the best of our knowledge, goroutines cannot be suspended through a public API; only the runtime can perform this operation, making it impractical to replicate this behavior in user code without additional mechanisms.

Java employs a similar short-path-slow-path idea, but its slow path is implemented using a modified version of the CLH \cite{Craig1993/CLH} algorithm. The primary waiting strategy relies on suspending (parking) the thread until it is explicitly unparked or a timeout occurs. If, after being unparked, progress still cannot be made, the thread performs a short active waiting phase and parks again. Note that the parking is used for both thread types, system and lightweight (aka virtual), omitting the use of the yielding for the latter.

Finally, it is worth noting the internal structure of the standard mutex implementations in Argobots and Boost Fibers. Despite minor architectural differences, both follow a conceptually similar design: an external flag used as a fast path and a waitlist of suspended threads protected by a spinlock. Upon attempting to acquire the mutex, a thread first tries to set the flag, if this attempt fails, it acquires the spinlock, enqueues itself in the waitlist, and suspends execution until explicitly resumed.

\section{Adapting Mutexes to Lightweight Threads}
\label{sec:common-adpating-method}

\subsection{Locks Waiting Primitives}

Original studies typically focus on the algorithmic design of locks, omitting low-level implementation details. Most locking algorithms share a common structure: a thread repeatedly checks a condition while waiting to acquire the lock. The essential question is what the thread should do during this waiting phase.

\paragraph{OS Threads.} First, the body of the loop may be empty or contain non-optimizable no-op instructions, commonly referred to as \code{cpu\_relax}. In this case, the waiting is purely active.
Second, we may inform the scheduler that the OS thread is waiting and request a context switch via a call such as \code{sched\_yield}.
However, usually this call is not used in implementations since it just make a request that the scheduler can ignore.
Third, a thread may be put to sleep for a certain period of time, for instance using \code{nanosleep}. Finally, a mechanism often referred to as sleep-wakeup can be implemented. There is no single standard implementation, but a classic approach is to use the futex system call. However, a futex uses kernel-level queue and should be regarded as an independent low-level primitive. 

\paragraph{Lightweight Thread.} Now, let us examine how the waiting primitives discussed above can be mapped to lightweight threading systems.
We took two commonly used libraries in C\code{++}: Argobots~\cite{Seo2017/Argobots} and Boost Fibers~\cite{BoostFiber}. Our goal is to identify direct analogues to OS thread waiting primitives in the libraries APIs.

Argobots is an execution system that offers the richest and, thus, low-level, set of tools for managing the lifecycle of lightweight threads. Therefore, in addition to context switching via \code{yield}, this model provides an explicit pair of commands, suspend and resume, which block a thread from scheduling and then put it back. However, the user must carefully monitor the order of suspend and resume calls: invoking resume before suspend will result in the thread sleeping indefinitely. For example, in Java, a suspend call may be ignored if a wakeup call was already performed. At the same time, the library does not provide timer-based sleeping, as the developers did not consider such functionality essential~\cite{ArgobotsSleepIssue}.

Boost Fibers is a library for working with lightweight threads that closely mirrors the interface of the \code{std::thread}. In addition to the standard \code{yield} operation, Boost Fibers also provides mechanisms for suspending fibers. This can be achieved through high-level abstractions such as \code{promise} or \code{condition\_variable}, or via the low-level scheduler API, where a fiber can be explicitly suspended and later resumed using a previously stored handle. In contrast to Argobots, Boost Fibers provide timer-based sleep methods, such as \code{sleep\_for}. 

Based on the surveyed capabilities, we will focus on two primary mechanisms: context switching and the \code{suspend-resume}. Timer-based sleep is not universally supported and, for short durations, may be nearly equivalent to a \code{yield}.

\subsection{Lock Implementations}
Summarizing the approaches to lock design for OS-level threads and other languages, three general waiting strategies can be identified. The first is active spinning, which is effective when the lock is expected to be released soon, but wastes CPU cycles if contention persists. The second involves an explicit yield, allowing other threads to progress while temporarily relinquishing control. Finally, suspending and explicitly resuming a thread minimizes CPU usage and frees lightweight thread pools, when the lock is held for a long time, avoiding unproductive spinning. Each strategy incurs higher overhead in that order but provides greater potential benefits. Therefore, combining these mechanisms in a balanced way, leveraging their strengths while mitigating their weaknesses, can yield more efficient lock implementations.

The questions whether all three parts are necessary and how they should be combined remain open. We will try to answer the first question in the experimental section. However, by default, we use the most balanced approach, which involves a spin–yield–suspend sequence, coordinated through a backoff mechanism. In this scheme, the longer a thread waits, the heavier the waiting strategy it employs. The key idea is that the time spent at each stage should be smaller than the overhead spent on the next threading mechanism. For example, if spinning takes longer than performing a yield and regaining control, the algorithm should use yield rather than making active spinning. According to our observations, the specific choice of backoff function is not crucial. We prefer exponential backoff, although other functions may be equally justified depending on the application context.

\subsubsection{MCS adaptation}
Let us illustrate this approach using the MCS lock, as it serves as a fundamental building block for more complex synchronization algorithms. The adaptation for TTAS would be identical, except that it does not involve thread suspension.

\begin{lstlisting}[float=t,caption={MCS modification},label=lst:mcs-modification]
void lock(LockNode node):
    predecessor = EXCHANGE(tail, node)
    if predecessor != nil:
        node.locked = true
        predecessor.next = node
        backoffPolicy = BackoffPolicy(node)
        while node.locked: #\label{line:mcs:loop1}#
            backoffPolicy.OnSpinWait()

void unlock(LockNode node):
    if node.next == nil
        if CAS(tail, node, nil): return
        backoffPolicy = BackoffPolicy(nil)
        while node.next == nil: #\label{line:mcs:loop2}#
            backoffPolicy.OnSpinWait()
    node.next.locked = false
    BackoffPolicy::Resume(node.next)
\end{lstlisting}

Recall that to acquire an MCS lock, each thread creates a node and passes it to the \code{lock} method. The nodes form a queue: the first node represents the lock owner, while all subsequent threads wait on a local flag within their node until the preceding thread updates it, thereby passing ownership of the lock.

Formally, the MCS algorithm contains several spin-waiting loops that can be modified. The first occurs in the \code{lock} method presented in Listing~\ref{lst:mcs-modification}, line~\ref{line:mcs:loop1}. When a thread attempting to acquire the lock enqueues itself, it begins spinning on its local flag while waiting for the predecessor thread to update it. This point is well-suited for integrating the three-stage waiting mechanism: depending on the queue length, the thread may benefit from actively waiting, yielding its context, or suspending.
Another spin loop is located in the \code{unlock} method, line~\ref{line:mcs:loop2}. It arises in cases where the successor has already enqueued itself but has not yet updated the predecessor's pointer. It is expected to be resolved within a very short time; therefore, suspension is unnecessary and may even be detrimental. Nevertheless, for safety, a backoff combined with context switching should still be applied. 

\begin{lstlisting}[float=t,caption={BackoffPolicy methods},label=lst:general-backoff]
class BackoffPolicy:
  Node node = nil
  int iterations = 0

  void OnSpinWait():
    iterations++
    if iterations < YIELD_LIMIT:
      // this time we do spin
      Spin(min(1 << iterations, SPIN_LIMIT))
    elif iterations < SUSPEND_LIMIT or node == nil:
      // we spin enough to amortize over yield
      Yield()
    else:
      // we had enough waiting to amortize over suspend
      TrySuspend(node) 

  static void Spin(ops):
    for (int i = 0; i < ops; i++) NOP
    
  static void TrySuspend(LockNode node):
    if CAS(node.resumeHandle, READY_FOR_SUSPEND, CreateResumeHandle(node)):
      InvokeLibrarySuspendLogic(resumeHandle) 

  static void Resume(LockNode node):
    resumeHandle = EXCHANGE(node.resumeHandle, KEEP_ACTIVE)
    if resumeHandle > KEEP_ACTIVE:
      InvokeLibraryResumeLogic(resumeHandle)        
\end{lstlisting}

The waiting mechanism is presented in Listing~\ref{lst:general-backoff}. To use this mechanism, an instance should be created before entering the loop, initialized either with a node containing the \code{resumeHandle} field for storing suspension data in the case of MCS, or with \code{nil} if suspension is not required in the case of TTAS. In each iteration of the spin-wait loop, the \code{OnSpinWait} method is invoked. Its logic is straightforward: the current iteration number is compared against two constants, \code{YIELD\_LIMIT} and \code{SUSPEND\_LIMIT}, which determine one of three waiting strategies. Initially, active spinning is used, followed by yielding, and finally, thread suspension when possible. To provide \code{suspend-resume} functionality, we use the \code{atomic\textless uintptr\_t\textgreater resumeHandle} field which stores the suspension data, typically the handle of a lightweight thread waiting on this node. Also, we reserve two special values \code{READY\_FOR\_SUSPEND = 0} and \code{KEEP\_ACTIVE = 1} in that field.

Now, we will briefly explain how \code{TrySuspend} and \code{Resume} methods work. To suspend, a thread atomically replaces the value \code{0} with a pointer to a control structure containing the necessary state for resumption. When another thread is responsible for waking the suspended thread, it first atomically replace the pointer with the value \code{1}. It then examines the previous value: if a valid pointer is observed, the waking thread is obligated to resume the suspended lightweight thread. This design enables safe and lock-free coordination between suspension and resumption, leveraging atomic operations to avoid race conditions while remaining compatible with any queue-lock structure.

\subsubsection{TTAS-MCS cohort lock}
Although we treat TTAS and MCS as the fundamental baseline primitives, the \ref{sec:evaluations} section will show that the relative performance of these locks has a great variance: on some settings, TTAS outperforms MCS significantly, and on some other settings, MCS outperforms TTAS.
To achieve the best of both worlds, we propose combining them via the cohort technique: TTAS-MCS-N lock.

This lock consists of two levels: an outer and an inner layer. The outer layer is implemented as an atomic boolean flag, while the inner layer comprises of $N$ MCS queues. Ownership of the lock requires holding only the outer flag. A thread attempting to acquire the lock first performs a single try-lock operation on the flag as a fast path. If this attempt fails, the thread joins one of the inner queues. The queue selection can be performed either by taking the current core ID modulo the number of queues, when the number of queues divides the number of cores, or by selecting a queue at random. In our observations, these approaches do not lead to significant differences in performance.

Once a queue is selected, a thread follows an MCS acquisition algorithm. When a thread becomes the head (owner) of its MCS queue, it competes with the heads of other queues by repeatedly attempting to acquire the outer flag in a TTAS-like loop. Each iteration of this loop employs the configured waiting mechanisms, except for thread suspension, which is not used for TTAS. When releasing the lock, a thread first clears the outer flag and then passes the ownership within its queue.

Note that the version of this lock, TTAS-MCS-1, is usually the main choice when writing the code in Java and is known as unfair ReentrantLock.
        
\section{Evaluation settings}
\label{sec:settings}

\paragraph{Benchmark structure.} To evaluate the performance of algorithms, usually it is preferred to use existing tools. However, well-known frameworks such as Google Benchmark~\cite{GoogleBenchmark} or Catch2~\cite{Catch2} were developed with the implicit assumption that the code will run on OS threads. Thus, we had to implement our own custom tool~\cite{NaslTools}.

\begin{lstlisting}[caption={Benchmark main testing loop},label=lst:common-test-loop]
while startTime + testTime < now():
    LOCK(mutex)
    CriticalSection()
    UNLOCK(mutex)
    ParallelWork()
\end{lstlisting}

We chose a standard mutex testing pattern for OS threads shown in Listing~\ref{lst:common-test-loop}. Over a limited time interval, at each step of the loop, a thread attempts to acquire the lock. Upon obtaining the lock, it executes the critical section code, releases the lock, and then performs actions that do not require synchronization. To highlight behavioural characteristics and compare mutex implementations, the main parameters of the scenario are \code{critical\_section} and \code{parallel\_work} functions. These two functions encapsulate arbitrary actions that are tested in combination with the mutex. 

We identify two metrics that are typically measured: throughput and latency. Throughput is calculated as the ratio of the number of successfully acquired locks to the total test time in seconds. The number of lock acquisitions is counted separately for each thread and then summed up. To avoid significant thread desynchronization, a barrier adapted for lightweight threads is placed before and after the testing loop. Lock latency is measured by recording timestamps immediately before and after \code{LOCK} function call, storing their differences, and computing quantiles (e.g., 0.95, 0.99) after the test.

\paragraph{Test scenarios.} For testing locks, we propose two fundamental scenarios: one with a short critical section and another with a long critical sections, both including context switches.
We base the first scenario on the one frequently used in lock studies, e.g., in~\cite{Dice2012/LockCohorting}. In this scenario, the critical section accesses two cache line aligned structures containing four integers each, increments every field once, and performs a context switch before exit. Including a switch inside the critical section highlights both the absence of busy-waiting issues and the latency of returning control to the lock owner, indirectly exposing lock overhead.
We chose the parallel section to perform $100$ iterations of $1000$ no-op instructions followed by a \code{yield}. It consists of multiple yield blocks to emulate realistic context-switch patterns. 

The second scenario reflects a behaviour that, to the best of our knowledge, has not been previously explored. This scenario emulates the call to some parallelisable function inside the critical section. For example, such cooperative behaviour was expected in~\cite{OpenBLASIssue2255}. in the critical section, $12$ threads are created to simulate parallel loop. Each of these threads executes $10\,000$ no-op instructions. The thread holding the lock joins all spawned threads before releasing the lock to ensure their completion. The parallel section consists of a 10-iteration loop, with each iteration executing $1000$ no-op instructions followed by a \code{yield}. An efficient mutex implementation in this scenario should promptly release processor resources from threads waiting for the lock, thereby enabling lightweight threads inside the critical section to proceed. The higher the efficiency of thread scheduling and lock handoff, the greater the achieved throughput. 

\section{Evaluation}
\label{sec:evaluations}

Doing experiments, we found that the performance of lock primitives in lightweight threading environments can differ significantly from that in OS-thread environments and often depends more on the underlying library than on the lock’s design itself.
So, instead of tuning locks for each separate scenario we try to answer two more general question.
First, we aim to determine whether it is necessary to use both sleep and yield operations, or if one of them is sufficient. We seek to identify cases where each of the discussed waiting strategies makes a meaningful difference. Second, as we found that TTAS and MCS can work differently in various scenarios, we examine whether we can achieve a relatively good performance in general using the cohort lock that unifies both locks, TTAS-MCS-N. 

In the previous section, we introduced two benchmark scenarios that form the basis of our experimental evaluation. For clarity, we refer to the scenario with a critical section using nested parallelism as ``Parallelizable CS'', and to the scenario with a critical section that increments two cache lines as ``Cache line increment CS''. These names will be used consistently throughout the section.

For most of our experiments, we use a system equipped with four Intel Xeon Gold 5218 CPUs, each containing 16 cores. Our focus is primarily on configurations with a limited number of active cores and a large number of lightweight threads. This setup reflects modern cloud environments, where minimizing the number of cores utilized by an instance simplifies its placement on arbitrary physical hosts. When a test involves more than 16 cores, they are allocated sequentially across NUMA nodes to ensure consistent and reproducible behavior.

Each benchmark consists of a series of identical runs. In each run, a specific mutex implementation is evaluated with a fixed number of lightweight threads and a fixed number of OS threads equal to the number of available cores. For clarity, the number of active cores, usually chosen as a power of two, is indicated on the plots by a vertical dashed line. Each run lasts for 10 seconds and is repeated 50 times, preceded by a warm-up phase of approximately 3 seconds. Each data point on the plots represents the median value across the 50 corresponding runs.

\begin{figure}[h]
    \begin{subfigure}{0.5\textwidth}
        \centering
        \includegraphics[width=\textwidth]{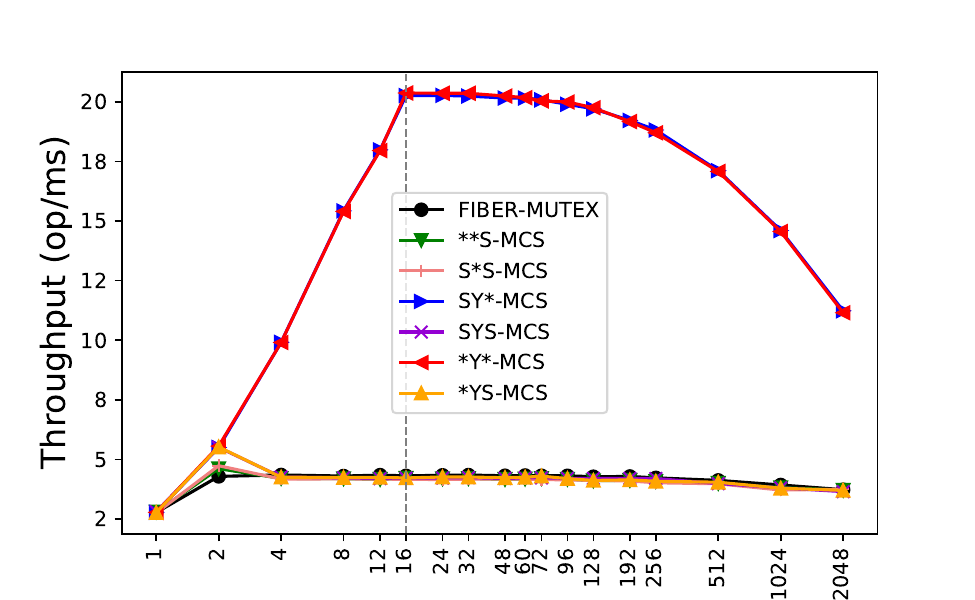}
        \caption{Parallelizable CS throughput}
        \label{fig:xeon-fibers-sys-mcs-16-pf-thr}
    \end{subfigure}\hfill
    \begin{subfigure}{0.5\textwidth}
        \centering
        \includegraphics[width=\textwidth]{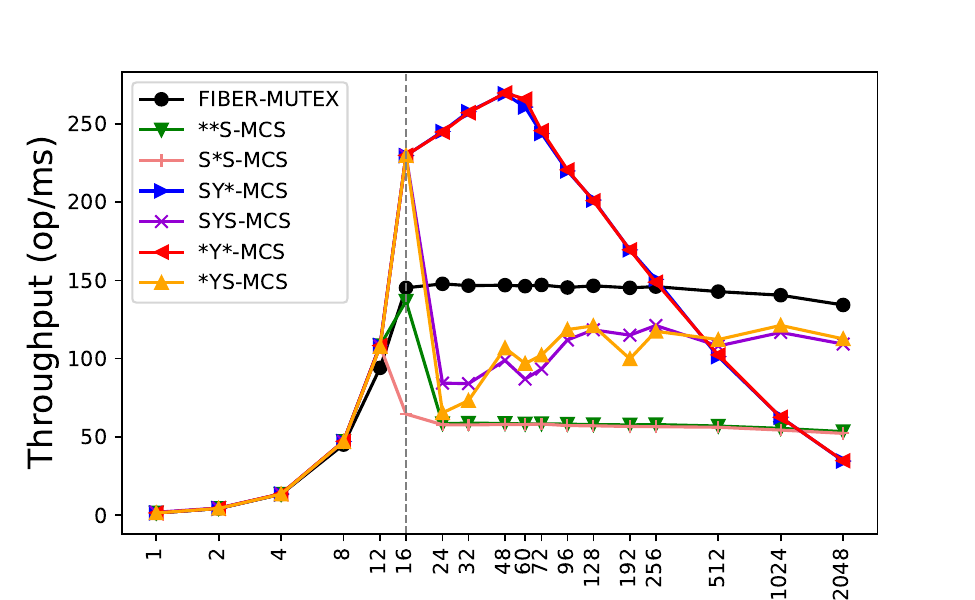}
        \caption{Cache line increment CS throughput}
        \label{fig:xeon-fibers-sys-mcs-16-2cl-inc-thr}
    \end{subfigure}
    \vspace{0.5cm}
    \begin{subfigure}{0.5\textwidth}
        \centering
        \includegraphics[width=\textwidth]{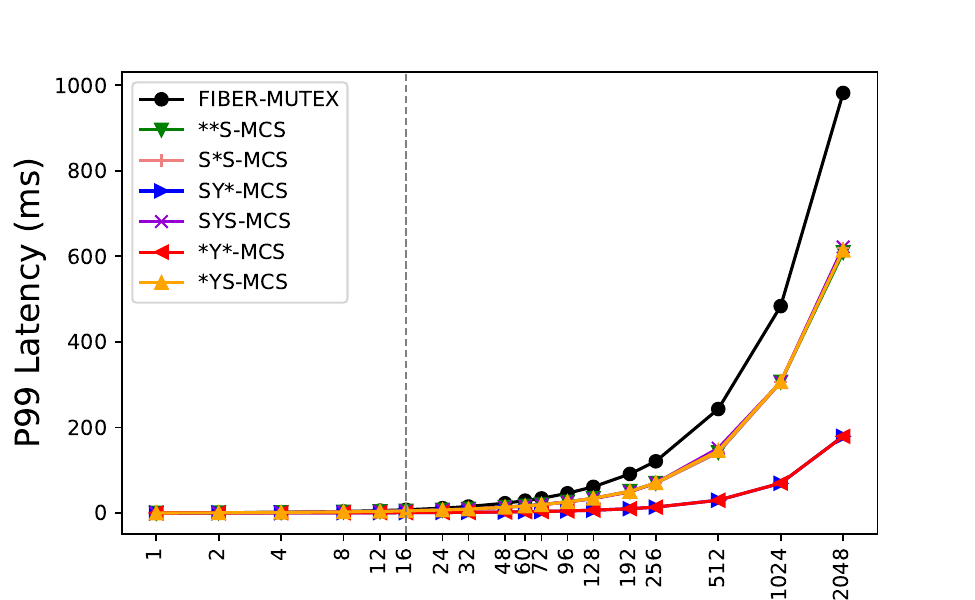}
        \caption{Parallelizable CS latency}
        \label{fig:xeon-fibers-sys-mcs-16-pf-lat}
    \end{subfigure}\hfill
    \begin{subfigure}{0.5\textwidth}
        \centering
        \includegraphics[width=\textwidth]{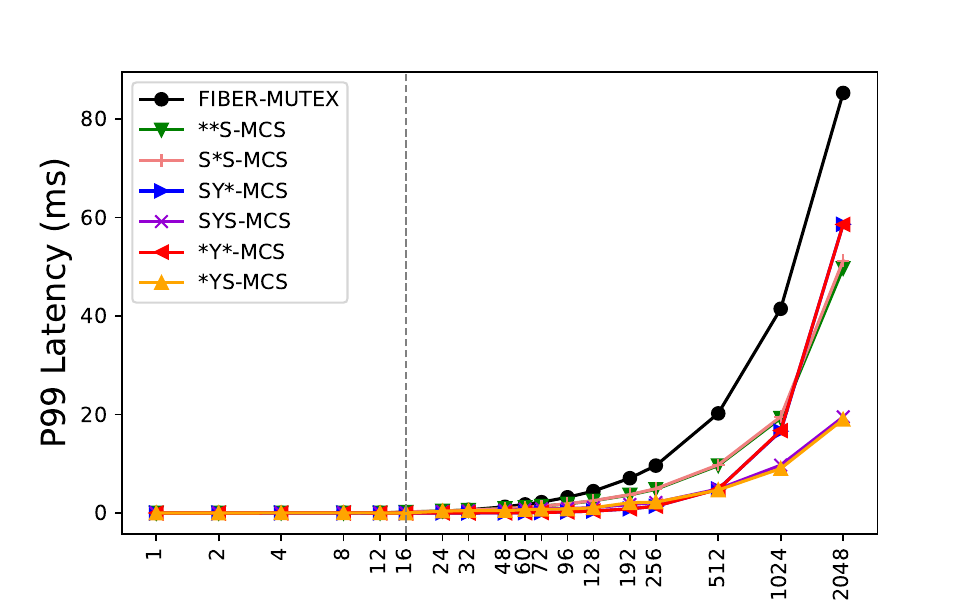}
        \caption{Cache line increment CS latency}
        \label{fig:xeon-fibers-sys-mcs-16-2cl-inc-lat}
    \end{subfigure}
    \caption{Waiting strategies comparison for Boost Fibers}
    \label{fig:xeon-fibers-sys-mcs-16}
\end{figure}

\subsection{Impact of Waiting Strategies on MCS Lock.}

We consider the simplest synchronization primitive that allows thread suspension, the MCS lock, evaluated under the two scenarios described earlier, but using different waiting techniques. The waiting strategies are always applied sequentially, as discussed in the Backoff section, although some of them may be disabled. For convenience, we introduce a three-letter notation: the first letter \textbf{S} stands for spin (active waiting), the second letter \textbf{Y} denotes the use of yield, and the third letter \textbf{S} represents suspend. If any of these letters is replaced by an asterisk (\textbf{*}), it indicates that the corresponding mechanism is disabled. For this section, we used a single NUMA node with $16$ cores, which we consider a reasonable configuration for an application instance while eliminating the influence of NUMA effects on the measurements.

First, we consider the results obtained for the Boost Fibers library in Figure~\labelcref{fig:xeon-fibers-sys-mcs-16}. While the plots contain many details, we focus only on the most notable observations, specifically, the cost of the yield and suspend operations. As shown in Figure~\labelcref{fig:xeon-fibers-sys-mcs-16-pf-thr}, the strategy based on \code{yield} alone, without suspension, significantly outperforms all others, including the default implementation (FIBER-MUTEX). This behavior can be explained by the fact that, up to $16$ threads, each lightweight thread effectively occupies its own core and competes only with additional threads inside the critical section. Consequently, yielding incurs almost no overhead, whereas primitives involving suspension spend extra time on thread parking without providing any advantage in this regime. However, as the number of threads grows, the cost of \code{yield} increases, and this strategy eventually becomes less efficient.

\begin{figure}[t]
    \begin{subfigure}[t]{0.5\textwidth}
        \centering
        \includegraphics[width=\textwidth]{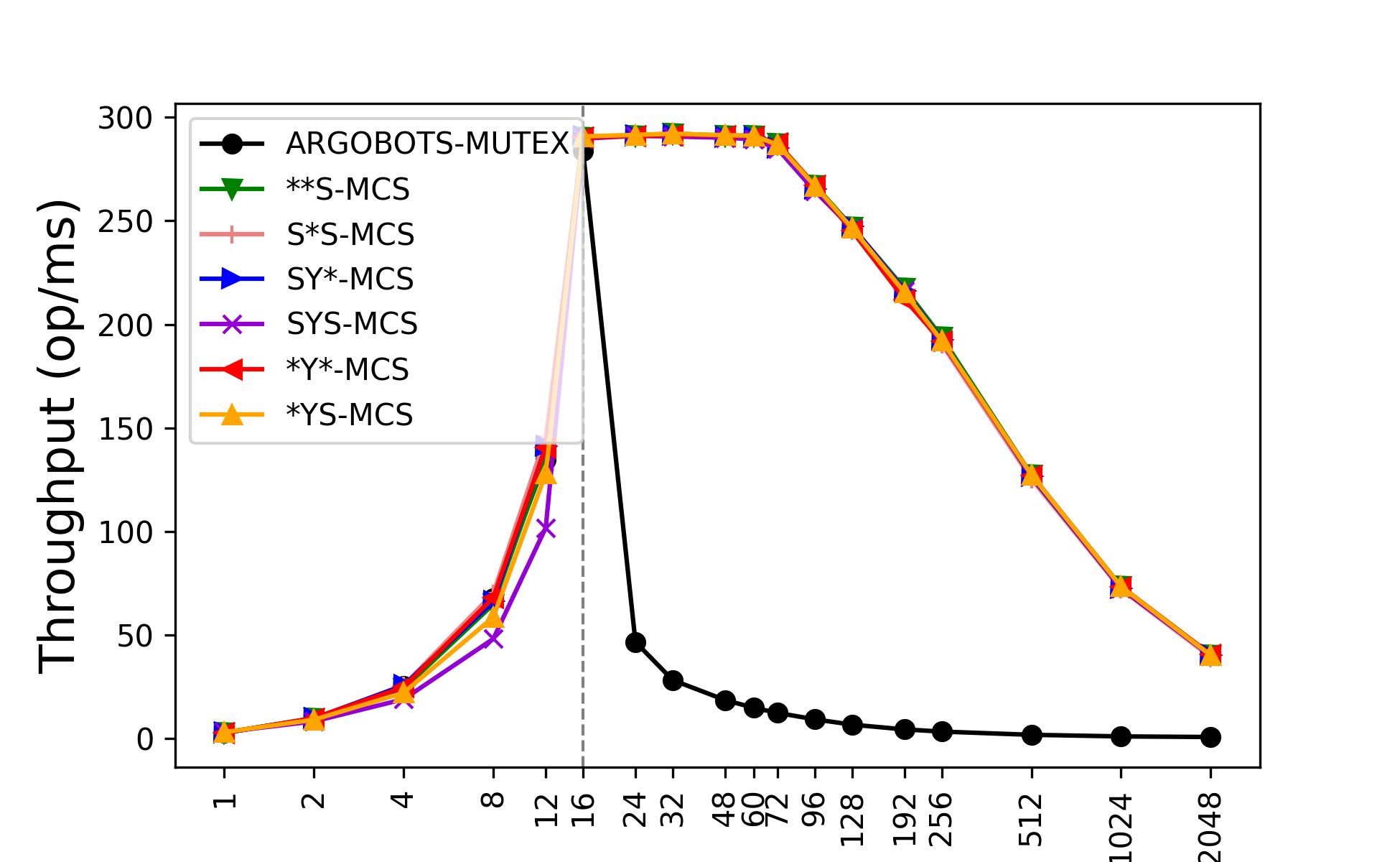}
        \caption{Cache line increment CS throughput}
        \label{fig:xeon-abt-sys-mcs-16-2-cl-inc-thr}
    \end{subfigure}\hfill
    \begin{subfigure}[t]{0.5\textwidth}
        \centering
        \includegraphics[width=\textwidth]{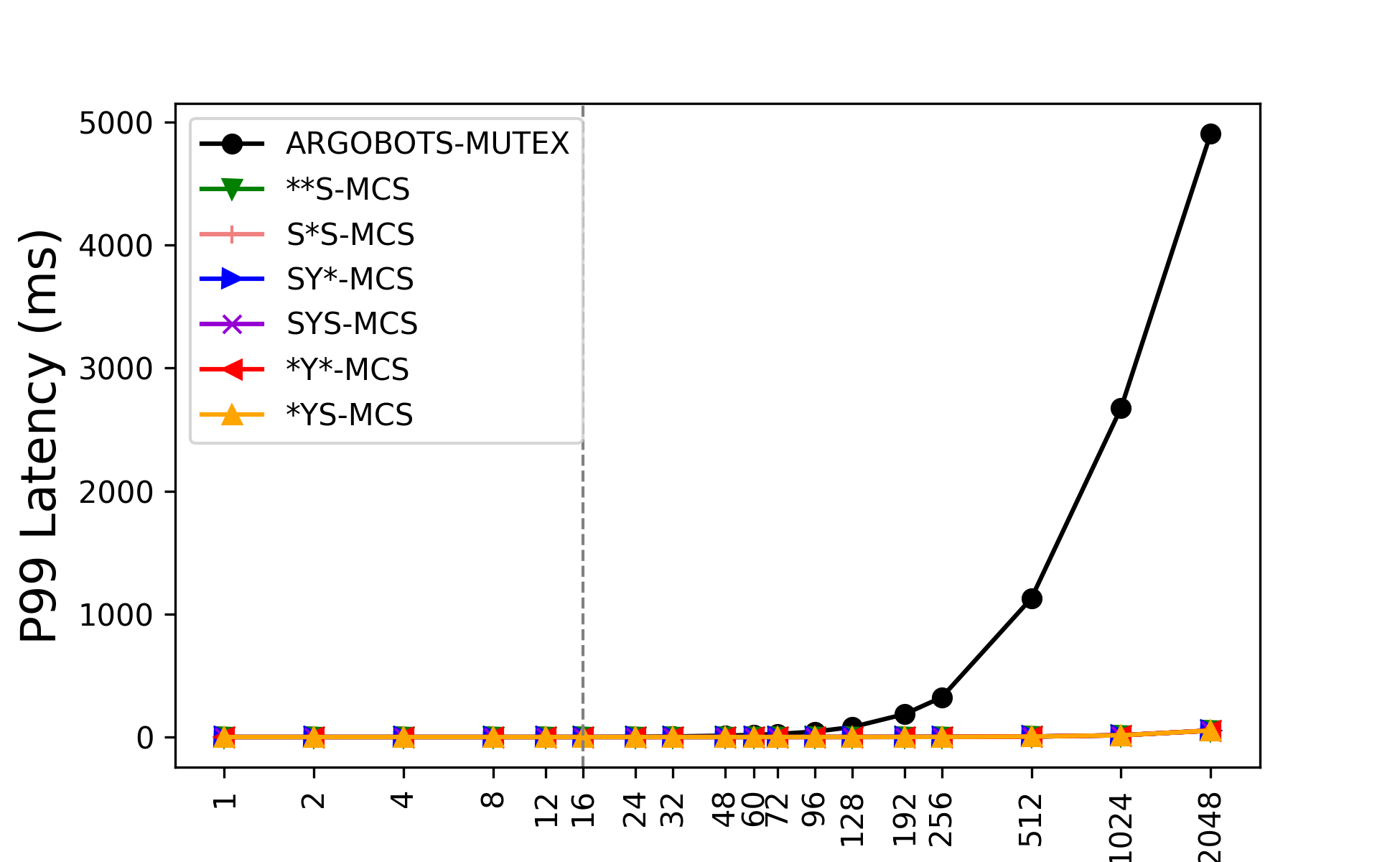}
        \caption{Cache line increment CS latency}
        \label{fig:xeon-abt-sys-mcs-16-2-cl-inc-lat}
    \end{subfigure}
    \caption{Waiting strategies comparison for Argobots}
    \label{fig:xeon-abt-sys-mcs-16-2-cl-inc}
\end{figure}

\begin{figure}[!t]
    \begin{subfigure}[t]{0.3\textwidth}
        \centering
        \includegraphics[width=\textwidth]{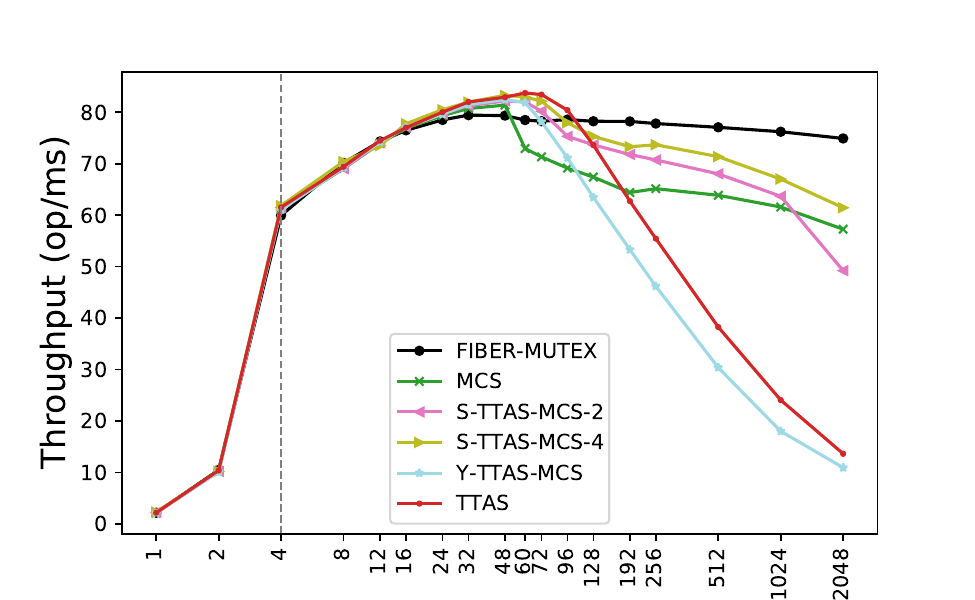}
        \caption{4 cores}
        \label{fig:xeon-fibers-ttas-mcs-4-2-cl-inc-thr}
    \end{subfigure}\hfill
    \begin{subfigure}[t]{0.3\textwidth}
        \centering
        \includegraphics[width=\textwidth]{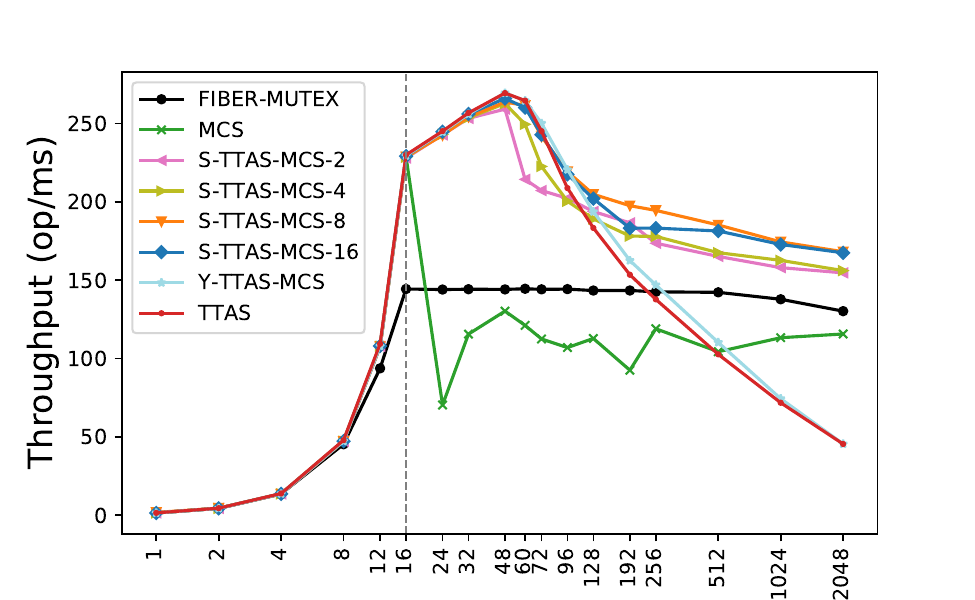}
        \caption{16 cores}
        \label{fig:xeon-fibers-ttas-mcs-16-2-cl-inc-thr}
    \end{subfigure}\hfill
    \begin{subfigure}[t]{0.3\textwidth}
        \centering
        \includegraphics[width=\textwidth]{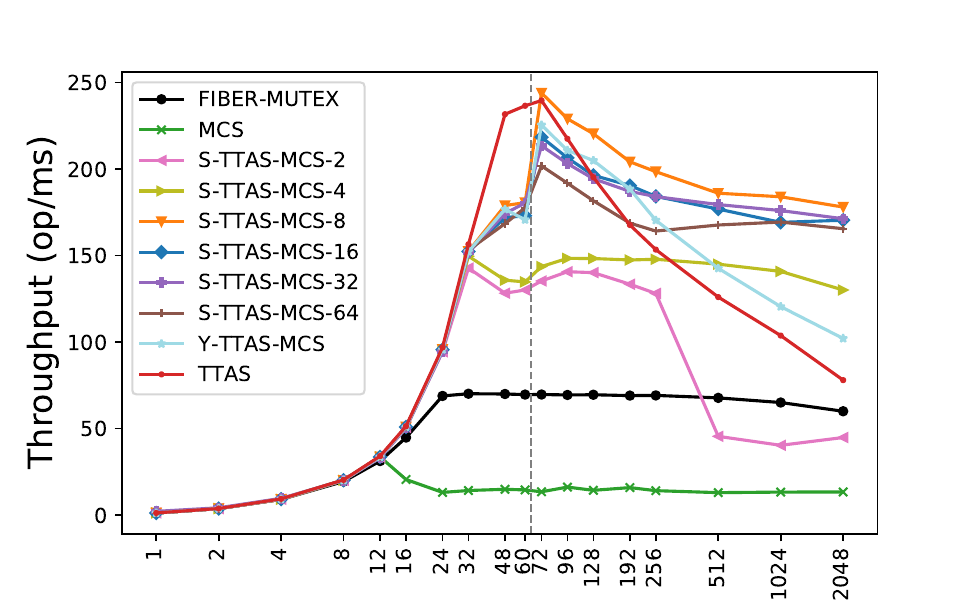}
        \caption{64 cores}
        \label{fig:xeon-fibers-ttas-mcs-64-2-cl-inc-thr}
    \end{subfigure}
    \caption{Cache line increment CS throughput for Boost Fibers}
    \label{fig:xeon-fibers-ttas-mcs-2-cl-inc-thr}
\end{figure}

\begin{figure}[!t]
    \begin{subfigure}[t]{0.3\textwidth}
        \centering
        \includegraphics[width=\textwidth]{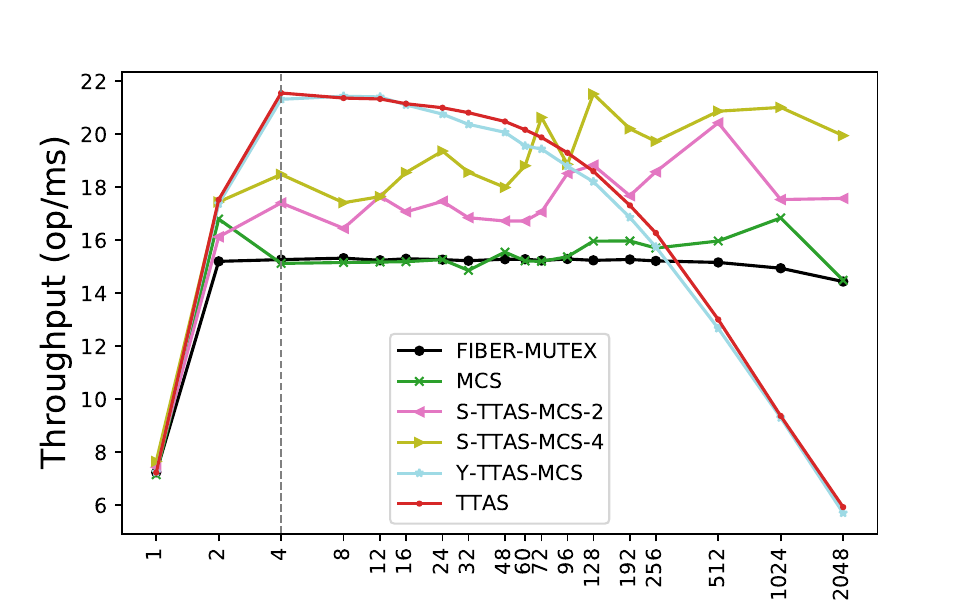}
        \caption{4 cores}
        \label{fig:xeon-fibers-ttas-mcs-4-pf-thr}
    \end{subfigure}\hfill
    \begin{subfigure}[t]{0.3\textwidth}
        \centering
        \includegraphics[width=\textwidth]{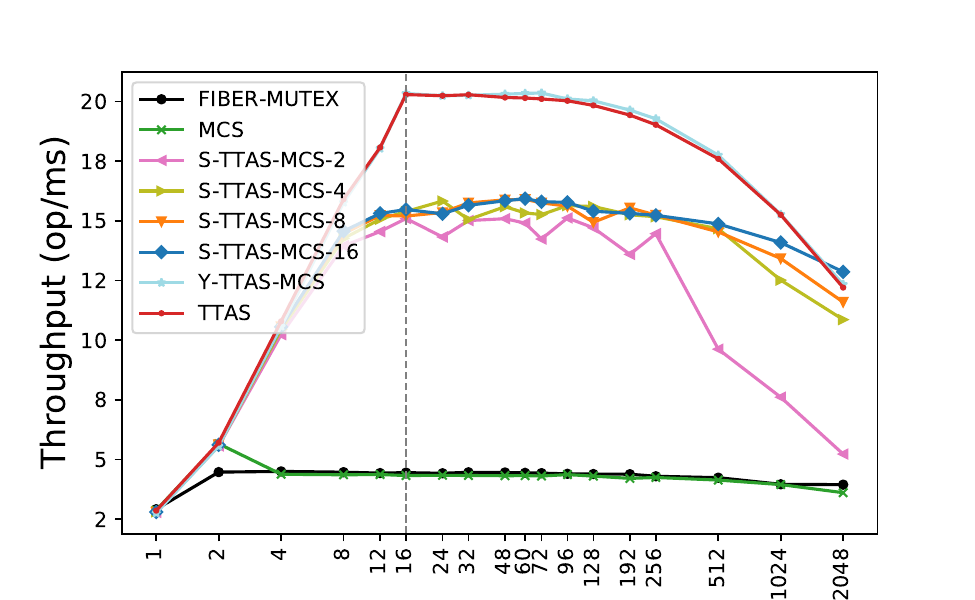}
        \caption{16 cores}
        \label{fig:xeon-fibers-ttas-mcs-16-pf-thr}
    \end{subfigure}\hfill
    \begin{subfigure}[t]{0.3\textwidth}
        \centering
        \includegraphics[width=\textwidth]{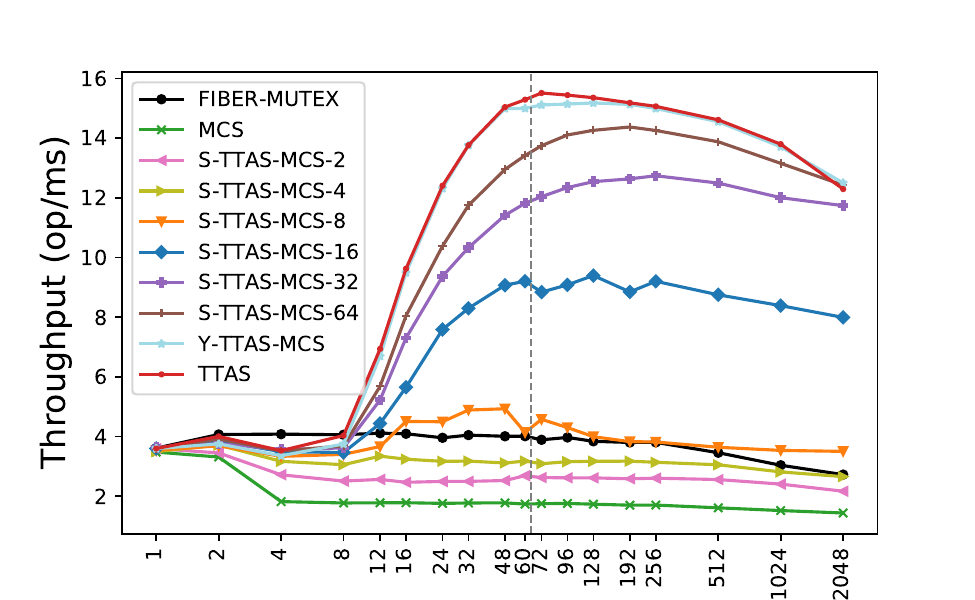}
        \caption{64 cores}
        \label{fig:xeon-fibers-ttas-mcs-64-pf-thr}
    \end{subfigure}
    \caption{Parallelizable CS throughput for Boost Fibers}
    \label{fig:xeon-fibers-ttas-mcs-pf-thr}
\end{figure}

\begin{figure}[t]
    \begin{subfigure}[t]{0.3\textwidth}
        \centering
        \includegraphics[width=\textwidth]{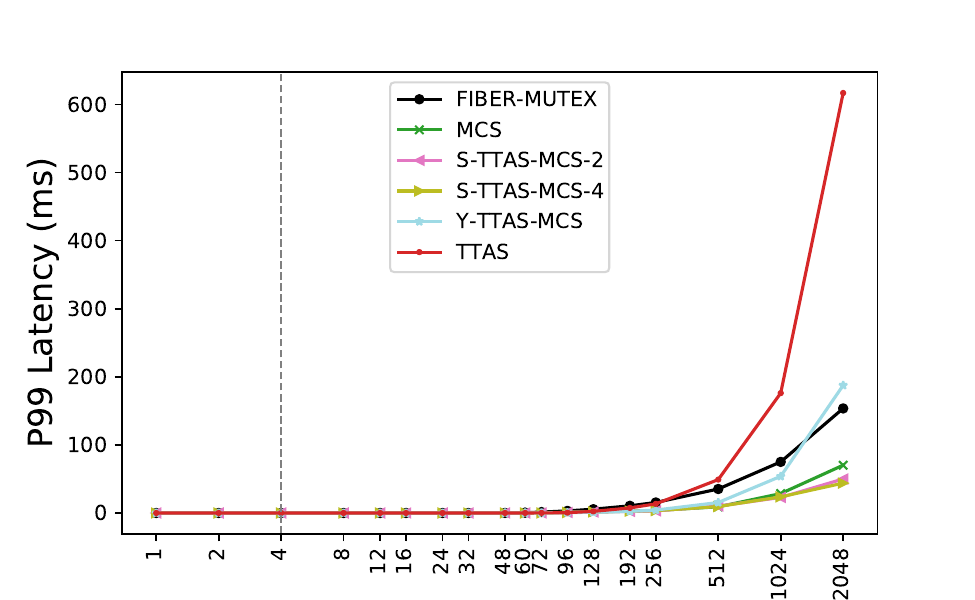}
        \caption{4 cores}
        \label{fig:xeon-fibers-ttas-mcs-4-2-cl-inc-lat}
    \end{subfigure}\hfill
    \begin{subfigure}[t]{0.3\textwidth}
        \centering
        \includegraphics[width=\textwidth]{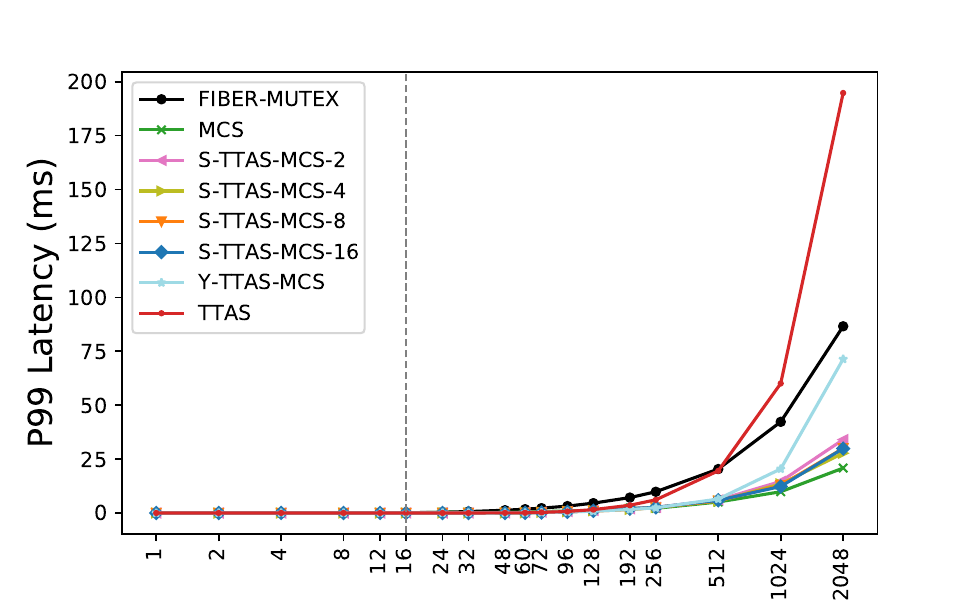}
        \caption{16 cores}
        \label{fig:xeon-fibers-ttas-mcs-16-2-cl-inc-lat}
    \end{subfigure}\hfill
    \begin{subfigure}[t]{0.3\textwidth}
        \centering
        \includegraphics[width=\textwidth]{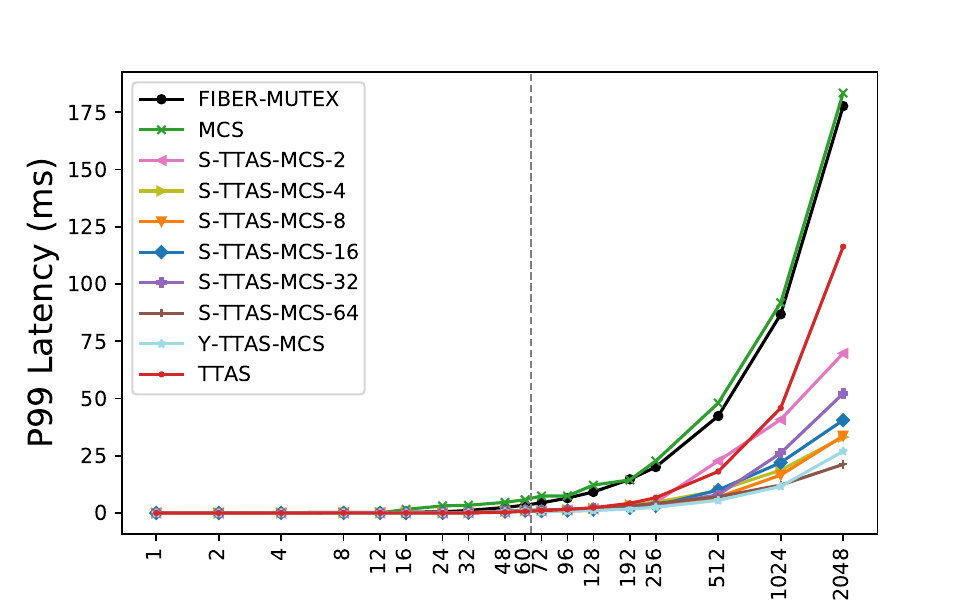}
        \caption{64 cores}
        \label{fig:xeon-fibers-ttas-mcs-64-2-cl-inc-lat}
    \end{subfigure}
    \caption{Cache line increment CS latency for Boost Fibers}
    \label{fig:xeon-fibers-ttas-mcs-2-cl-inc-lat}
\end{figure}

\begin{figure}
    \begin{subfigure}[t]{0.3\textwidth}
        \centering
        \includegraphics[width=\textwidth]{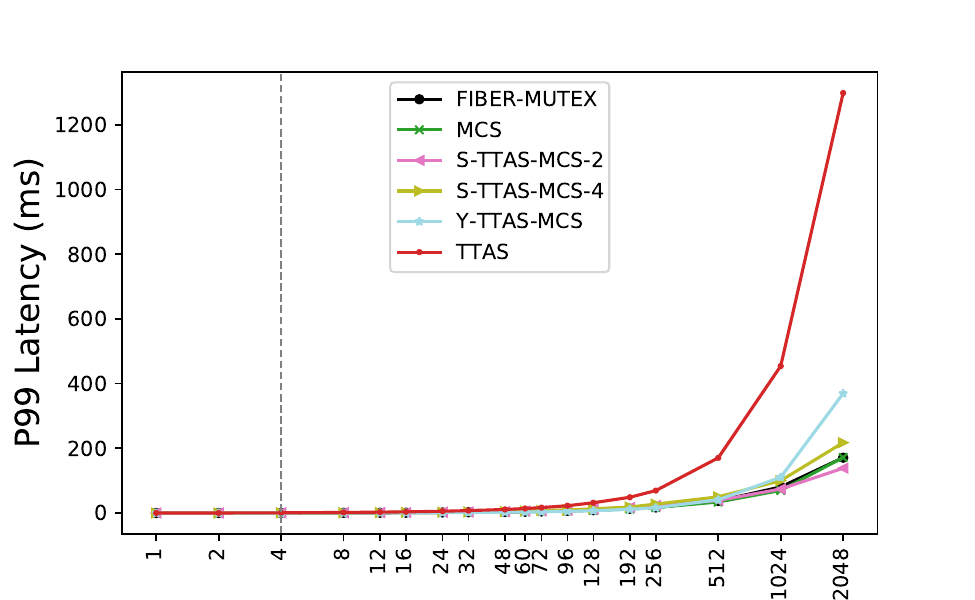}
        \caption{4 cores}
        \label{fig:xeon-fibers-ttas-mcs-4-pf-lat}
    \end{subfigure}\hfill
    \begin{subfigure}[t]{0.3\textwidth}
        \centering
        \includegraphics[width=\textwidth]{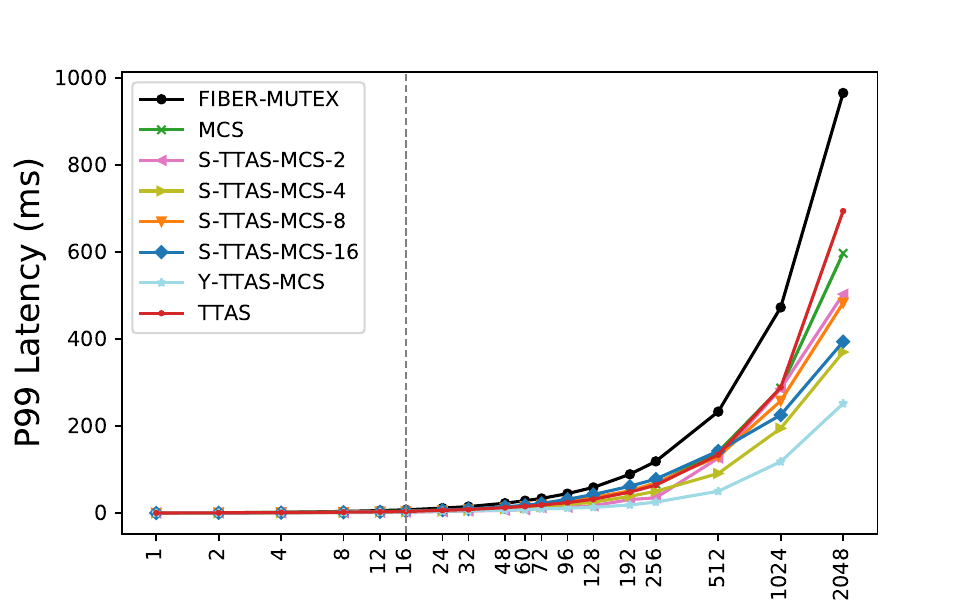}
        \caption{16 cores}
        \label{fig:xeon-fibers-ttas-mcs-16-pf-lat}
    \end{subfigure}\hfill
    \begin{subfigure}[t]{0.3\textwidth}
        \centering
        \includegraphics[width=\textwidth]{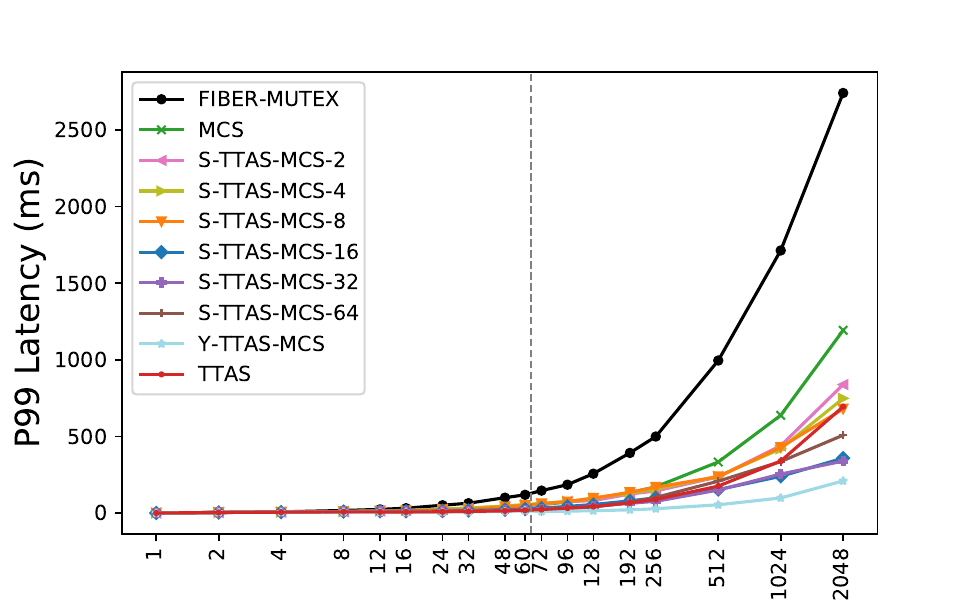}
        \caption{64 cores}
        \label{fig:xeon-fibers-ttas-mcs-64-pf-lat}
    \end{subfigure}
    \caption{Parallelizable CS latency for Boost Fibers}
    \label{fig:xeon-fibers-ttas-mcs-pf-lat}
\end{figure}

When examining throughput for a much shorter critical section, as shown in Figure~\labelcref{fig:xeon-fibers-sys-mcs-16-2cl-inc-thr}, several observations can be made. The strategy that combines yield and suspend performs comparably to the others under low contention and maintains stable results throughout the test. In contrast, the yield-only versions degrade significantly as the number of threads increases. At the same time, the library implementation, in which a thread checks an external flag and suspends if unsuccessful, performs slightly better than our version in this case, likely due to implementation details and parameter tuning.

The latency results in Figures~\labelcref{fig:xeon-fibers-sys-mcs-16-pf-lat,fig:xeon-fibers-sys-mcs-16-2cl-inc-lat} generally follow the same trends as throughput, except for the library lock, which consistently exhibits the worst performance. This confirms that immediate thread suspension is not an effective strategy in this context.

For Argobots, we present only the Cache line scenario in Figure~\labelcref{fig:xeon-abt-sys-mcs-16-2-cl-inc}, since all modifications produced nearly identical results in both cases. As we checked, the costs of yield and suspend in Argobots do not differ significantly.

\subsection{Queue Scaling Across Cores}
In this section, we examine how the number of CPU cores influences the optimal number of queues in the TTAS-MCS-N lock. This lock combines an outer flag as the fast path with a configurable number of internal MCS queues, whose owners compete for that flag. For comparison, we also include the standard library mutex and the individual TTAS and MCS components.
As before, we vary the waiting strategies but simplify the notation to two letters. Lock names starting with \textbf{Y} use active waiting and yield only, while those starting with \textbf{S} employ the full three-stage mechanism. The number of queues is explicitly indicated in each lock’s name. In cases where the differences between configurations with varying numbers of queues are negligible, only a single representative primitive is shown in the plots, for example, Y-TTAS-MCS.

For this experiment, we selected three core counts: $4$, $16$, and $64$.
The first configuration provides limited but noticeable concurrency, the second represents a balanced setup corresponding to a single NUMA node, and the last utilizes all available cores on the machine.

\begin{figure}[!t]
    \begin{subfigure}[t]{0.5\textwidth}
        \centering
        \includegraphics[width=\textwidth]{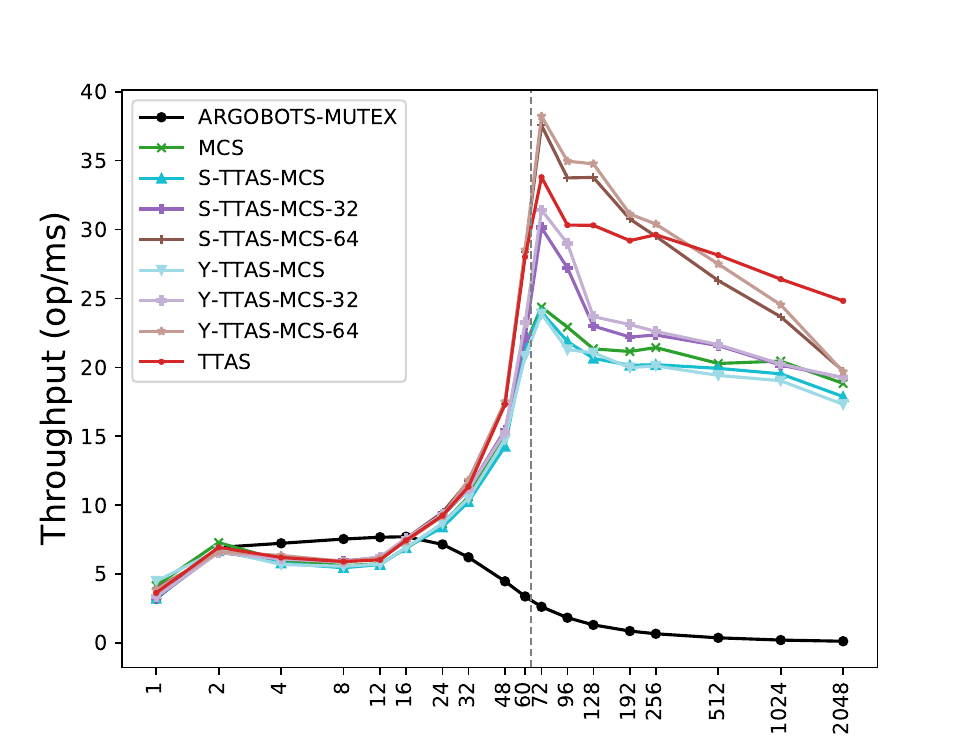}
        \caption{Parallelizable CS throughput}
        \label{fig:xeon-abt-ttas-mcs-64-pf-thr}
    \end{subfigure}\hfill
    \begin{subfigure}[t]{0.5\textwidth}
        \centering
        \includegraphics[width=\textwidth]{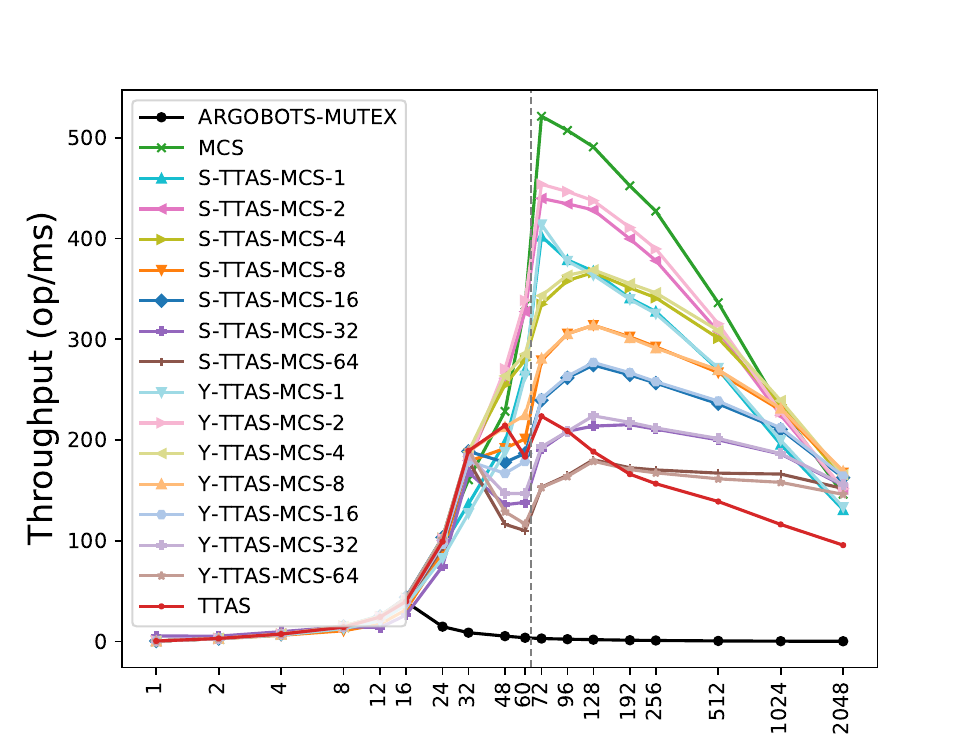}
        \caption{Cache line increment CS throughput}
        \label{fig:xeon-abt-ttas-mcs-64-2cl-inc-thr}
    \end{subfigure}
    \caption{Both CS throughput for Argobots (64 cores)}
    \label{fig:xeon-abt-ttas-mcs-64-thr}
\end{figure}  

As in the previous section, a variety of observations can be drawn from the results presented in Figures~\labelcref{fig:xeon-fibers-ttas-mcs-2-cl-inc-thr,fig:xeon-fibers-ttas-mcs-pf-thr,fig:xeon-fibers-ttas-mcs-2-cl-inc-lat,fig:xeon-fibers-ttas-mcs-pf-lat}. Here, we highlight only the most relevant ones. For short critical sections (Figures~\labelcref{fig:xeon-fibers-ttas-mcs-2-cl-inc-thr,fig:xeon-fibers-ttas-mcs-2-cl-inc-lat}), the configuration with eight queues, S-TTAS-MCS-8 (and four, S-TTAS-MCS-4, in the first case, since allocating more queues than cores provides no benefit), demonstrates near-optimal performance. This configuration achieves a favorable balance between throughput and latency. For longer critical sections and larger core counts (Figures~\labelcref{fig:xeon-fibers-ttas-mcs-pf-thr,fig:xeon-fibers-ttas-mcs-pf-lat}), it may be more reasonable to prefer primitives that rely solely on yield, such as Y-TTAS-MCS. This configuration combines the high throughput of TTAS with the low latency characteristic of fair queue-based locks. This trend is most evident in Figure~\labelcref{fig:xeon-fibers-ttas-mcs-64-pf-thr}, where throughput increases with the number of queues, and the results fall between the two extremes: MCS with suspend and MCS with yield only.

For Argobots, the results are far less intriguing. On 4- and 16-core configurations, the performance of all synchronization primitives is either identical to that in Figure~\labelcref{fig:xeon-abt-sys-mcs-16-2-cl-inc-thr} or close enough to draw the same conclusions, thus, these graphs are omitted. Likewise, latency measurements offer little additional value, as their visual representation is almost indistinguishable from Figure~\labelcref{fig:xeon-abt-sys-mcs-16-2-cl-inc-lat}. Even when excluding the library mutex, no meaningful insights can be derived. 

The most notable differences appear in both scenarios on 64 cores. In Figure~\labelcref{fig:xeon-abt-ttas-mcs-64-pf-thr}, we combined the cohort locks with fewer than 32 queues under the label S/Y-TTAS-MCS, since their variations fall within the margin of error. Although the results for the ``Parallelizable CS'' scenario differ from those of Boost Fibers, they follow a similar trend: modifications whose behavior most closely resembles TTAS demonstrate the highest throughput. 

For the ``Cache Line Increment CS'' scenario shown in Figure~\labelcref{fig:xeon-abt-ttas-mcs-64-2cl-inc-thr}, a clear dependence on the number of queues can be observed, while no such dependence is evident for the waiting mechanism. The trends differ from those seen in Boost Fibers more noticeably than in the previous figure, though they can still be regarded as broadly similar. Specifically, the best performance is achieved by the lock that strikes a balance between active waiting and thread suspension, effectively freeing up processor resources.  

\vspace{-0.2cm}
\section{Conclusion}
\label{sec:conclusion}
In this work, we examined two fundamental synchronization primitives, TTAS and MCS, as well as the most common technique for combining them through cohort-based locking as used in industry. We analysed the waiting mechanisms required for lock operation in systems with lightweight threads and explored their behavior across various configurations and combinations.

At first, the various waiting strategies demonstrated effects of comparable importance. Specifically, we observed that relying solely on \code{yield} in scenarios with long critical sections can provide a substantial performance improvement as the number of cores increases, across a wide range of lightweight thread counts. In contrast, for short critical sections, a weighted lock that employs all three waiting stages, including thread suspension, performs significantly better than the yield-based variants.

Then, after finding that MCS and TTAS can have different relative performances in varying scenarios, we tested the cohort lock, TTAS-MCS-N, which should strike a balance between TTAS and MCS depending on N.
We conducted a series of experiments with different core counts to support our claim.
In some scenarios, the cohort lock outperforms both original primitives across all metrics, while in others it provides advantages in specific metrics without sacrificing performance in the remaining ones. So, if one should choose some lock for an unknown setting, our proposal would be to choose this cohort lock.

Although the performance characteristics of Argobots and Boost Fibers differ, sometimes substantially, we believe that we have identified several common trends that may be useful to practitioners. In our projects, we have repeatedly encountered attempts to migrate code to both of these libraries. A recurring question in such efforts is whether it is sufficient to insert \code{yield} at all potential points of active waiting, particularly inside mutex implementations, or whether a more optimal approach exists that can deliver measurable performance gains. We hope that the results presented in this study provide, at least in part, an answer to this question and assist in more effectively tuning mutex behavior for these libraries.

Looking ahead, we plan to focus on determining the most effective configurations and parameter values for combining different waiting mechanisms. This should enable either compile-time specialization of mutexes for specific workloads or the development of an adaptive scheme capable of efficiently adjusting to any target library.



\bibliography{references.bib}

@article{Sutter2005/FreeLauncIsOver,
  title={The free lunch is over: A fundamental turn toward concurrency in software},
  author={Sutter, Herb and others},
  journal={Dr. Dobb’s journal},
  volume={30},
  number={3},
  pages={202--210},
  year={2005}
}

@article{Seo2017/Argobots,
  title={Argobots: A lightweight low-level threading and tasking framework},
  author={Seo, Sangmin and Amer, Abdelhalim and Balaji, Pavan and Bordage, Cyril and Bosilca, George and Brooks, Alex and Carns, Philip and Castell{\'o}, Adri{\'a}n and Genet, Damien and Herault, Thomas and others},
  journal={IEEE Transactions on Parallel and Distributed Systems},
  volume={29},
  number={3},
  pages={512--526},
  year={2017},
  publisher={IEEE}
}

@article{Castello2020/AnalysisLTLibs,
  title={Analysis of threading libraries for high performance computing},
  author={Castell{\'o}, Adri{\'a}n and Gual, Rafael Mayo and Seo, Sangmin and Balaji, Pavan and Quintana-Orti, Enrique S and Pena, Antonio J},
  journal={IEEE Transactions on Computers},
  volume={69},
  number={9},
  pages={1279--1292},
  year={2020},
  publisher={IEEE}
}

@article{MellorCrummey1991/MCS,
  author = {Mellor-Crummey, John M. and Scott, Michael L.},
  title = {Algorithms for scalable synchronization on shared-memory multiprocessors},
  year = {1991},
  issue_date = {Feb. 1991},
  publisher = {Association for Computing Machinery},
  address = {New York, NY, USA},
  volume = {9}, 
  number = {1},
  issn = {0734-2071},
  url = {https://doi.org/10.1145/103727.103729},
  doi = {10.1145/103727.103729},
  journal = {ACM Trans. Comput. Syst.},
  month = {feb},
  pages = {21–65},
  numpages = {45}
}

@article{Chabbi2016/HMCS,
  author = {Chabbi, Milind and Fagan, Michael and Mellor-Crummey, John},
  title = {High performance locks for multi-level {NUMA} systems},
  year = {2015},
  issue_date = {August 2015},
  publisher = {Association for Computing Machinery},
  address = {New York, NY, USA},
  volume = {50},
  number = {8},
  issn = {0362-1340},
  url = {https://doi.org/10.1145/2858788.2688503},
  doi = {10.1145/2858788.2688503},
  journal = {SIGPLAN Not.},
  month = {jan},
  pages = {215–226},
  numpages = {12}
}

@article{Lameter2013/NUMA,
  title={{NUMA} (Non-Uniform Memory Access): An Overview: {NUMA} becomes more common because memory controllers get close to execution units on microprocessors.},
  author={Lameter, Christoph},
  journal={Queue},
  volume={11},
  number={7},
  pages={40--51},
  year={2013},
  publisher={ACM New York, NY, USA}
}

@article{Goodacre2005/ARM,
  title={Parallelism and the {ARM} instruction set architecture},
  author={Goodacre, John and Sloss, Andrew N},
  journal={Computer},
  volume={38},
  number={7},
  pages={42--50},
  year={2005},
  publisher={IEEE}
}

@book{Tanenbaum2015/ModernOS,
  title={Modern operating systems},
  author={Tanenbaum, Andrew S and Bos, Herbert},
  year={2015},
  publisher={Pearson Education}
}

@book{Shanley2010/x86,
  title={x86 Instruction Set Architecture},
  author={Shanley, Tom},
  year={2010},
  publisher={MindShare press}
}

@book{Donovan2015/Go,
  title={The Go programming language},
  author={Donovan, Alan AA and Kernighan, Brian W},
  year={2015},
  publisher={Addison-Wesley Professional}
}

@inproceedings{Rudolph1984/TTAS,
  title={Dynamic decentralized cache schemes for {MIMD} parallel processors},
  author={Rudolph, Larry and Segall, Zary},
  booktitle={Proceedings of the 11th Annual International Symposium on Computer Architecture},
  pages={340--347},
  year={1984}
}

@inproceedings{Dice2019/CNA,
  author = {Dice, Dave and Kogan, Alex},
  title = {Compact {NUMA-aware} Locks},
  year = {2019},
  isbn = {9781450362818},
  publisher = {Association for Computing Machinery},
  address = {New York, NY, USA},
  url = {https://doi.org/10.1145/3302424.3303984},
  doi = {10.1145/3302424.3303984},
  booktitle = {Proceedings of the Fourteenth EuroSys Conference 2019},
  articleno = {12},
  numpages = {15},
  location = {Dresden, Germany},
  series = {EuroSys '19}
}

@inproceedings{Oberhauser2021/HMCSArm,
  author="Oberhauser, Jonas and Oberhauser, Lilith and Paolillo, Antonio and Behrens, Diogo and Fu, Ming and Vafeiadis, Viktor",
  editor="Echihabi, Karima and Meyer, Roland",
  title="Verifying and Optimizing the {HMCS} Lock for {ARM} Servers",
  booktitle="Networked Systems",
  year="2021",
  publisher="Springer International Publishing",
  address="Cham",
  pages="240--260",
  isbn="978-3-030-91014-3"
}

@inproceedings{Dice2012/LockCohorting,
    author = {Dice, David and Marathe, Virendra J. and Shavit, Nir},
    title = {Lock cohorting: a general technique for designing {NUMA} locks},
    year = {2012},
    isbn = {9781450311601},
    publisher = {Association for Computing Machinery},
    address = {New York, NY, USA},
    url = {https://doi.org/10.1145/2145816.2145848},
    doi = {10.1145/2145816.2145848},
    booktitle = {Proceedings of the 17th ACM SIGPLAN Symposium on Principles and Practice of Parallel Programming},
    pages = {247–256},
    numpages = {10},
    location = {New Orleans, Louisiana, USA},
    series = {PPoPP '12}
}

@inproceedings{Dice2025/ReciprocatingLocks,
    author = {Dice, Dave and Kogan, Alex},
    title = {Reciprocating Locks},
    year = {2025},
    isbn = {9798400714436},
    publisher = {Association for Computing Machinery},
    address = {New York, NY, USA},
    url = {https://doi.org/10.1145/3710848.3710862},
    doi = {10.1145/3710848.3710862},
    booktitle = {Proceedings of the 30th ACM SIGPLAN Annual Symposium on Principles and Practice of Parallel Programming},
    pages = {85–98},
    numpages = {14},
    location = {Las Vegas, NV, USA},
    series = {PPoPP '25}
}

@inproceedings{Shiina2021/PreemptiveLightThreads,
    author = {Shiina, Shumpei and Iwasaki, Shintaro and Taura, Kenjiro and Balaji, Pavan},
    title = {Lightweight preemptive user-level threads},
    year = {2021},
    isbn = {9781450382946},
    publisher = {Association for Computing Machinery},
    address = {New York, NY, USA},
    url = {https://doi.org/10.1145/3437801.3441610},
    doi = {10.1145/3437801.3441610},
    booktitle = {Proceedings of the 26th ACM SIGPLAN Symposium on Principles and Practice of Parallel Programming},
    pages = {374–388},
    numpages = {15},
    location = {Virtual Event, Republic of Korea},
    series = {PPoPP '21}
}

@inproceedings{Madsen2021/PreemptiveCoroutines,
    author = {Madsen, Ole Lehrmann},
    title = {Using Coroutines for Multi-core Preemptive Scheduling},
    year = {2021},
    isbn = {9781450387071},
    publisher = {Association for Computing Machinery},
    address = {New York, NY, USA},
    url = {https://doi.org/10.1145/3477113.3487271},
    doi = {10.1145/3477113.3487271},
    booktitle = {Proceedings of the 11th Workshop on Programming Languages and Operating Systems},
    pages = {46–52},
    numpages = {7},
    location = {Virtual Event, Germany},
    series = {PLOS '21}
}

@inproceedings{Boucher2020/LwPreemptibleFunc,
  title={Lightweight preemptible functions},
  author={Boucher, Sol and Kalia, Anuj and Andersen, David G and Kaminsky, Michael},
  booktitle={2020 USENIX Annual Technical Conference (USENIX ATC 20)},
  pages={465--477},
  year={2020}
}

@inproceedings{Mollison2013/UserspaceLibWithPreemptiveThreads,
  author={Mollison, Malcolm S. and Anderson, James H.},
  booktitle={2013 IEEE 19th Real-Time and Embedded Technology and Applications Symposium ({RTAS})}, 
  title={Bringing theory into practice: A userspace library for multicore real-time scheduling}, 
  year={2013},
  volume={},
  number={},
  pages={283-292},
  doi={10.1109/RTAS.2013.6531100}
}

@techreport{Craig1993/CLH,
  title={Building {FIFO} and priorityqueuing spin locks from atomic swap},
  author={Craig, Travis},
  year={1993},
  institution={Technical Report TR 93-02-02, Department of Computer Science, University of Washington}
}

@misc{Moore1965,
  title={Cramming more components onto integrated circuits},
  author={Moore, Gordon E and others},
  year={1965}
}

@online{BoostFiber,
  key = {GitHub Repository: Boost Fibers},
  title = {Boost Fiber library repository},
  url = {https://github.com/boostorg/fiber},
  urldate = {2025-07-22},
  langid = {english}
}

@online{GoogleBenchmark,
  key = {GitHub repository: Google benchmark},
  title = {Google benchmark},
  url = {https://github.com/google/benchmark},
  urldate = {2025-06-14},
  langid = {english}
}

@online{Catch2,
  key = {GitHub repository: Catch2},
  title = {Unit testing framework Catch2},
  url = {https://github.com/catchorg/Catch2},
  urldate = {2025-06-14},
  langid = {english}
}

@online{ArgobotsSleepIssue,
  key = {Argobots Sleep Issue},
  title = {Argobots Sleep Issue},
  url = {https://github.com/pmodels/argobots/issues/29},
  urldate = {2025-08-31},
  langid = {english}
}

@online{Golang114,
  key = {Go 1.14 Release Notes},
  title = {Go 1.14 Release Notes},
  url = {https://go.dev/doc/go1.14},
  urldate = {2025-09-06},
  langid = {english}
}

@online{JavaJep491,
  key = {JEP 491: Synchronize Virtual Threads without Pinning},
  title = {{JEP} 491: Synchronize Virtual Threads without Pinning},
  url = {https://openjdk.org/jeps/491},
  urldate = {2025-09-06},
  langid = {english}
}

@online{OpenBLASIssue2255,
  key = {OpenBLAS: replacing threading backend issue},
  title = {OpenBLAS: replacing threading backend issue},
  url = {https://github.com/OpenMathLib/OpenBLAS/pull/2255},
  urldate = {2025-11-07},
  langid = {english}
}

@online{NaslTools,
  key = {nasl-tools},
  title = {{NASL}-tools: benchmark and locks implementations repository},
  url = {https://github.com/tarasska/nasl-tools},
  urldate = {2025-12-09},
  langid = {english}
}

\end{document}